\documentclass[a4paper]{article}

\usepackage{INTERSPEECH2022}
\usepackage{threeparttable}
\usepackage{hyperref}

\title{NAS-VAD: Neural Architecture Search for Voice Activity Detection}
\name{Daniel Rho$^1$, Jinhyeok Park$^{1,2}$, and Jong Hwan Ko$^1$}
\address{
  $^1$Sungkyunkwan University \\
  $^2$Graduate School of AI, POSTECH}
\email{daniel231@skku.edu, jinhyeok1234@postech.ac.kr, jhko@skku.edu}

\begin{document}

\maketitle
\begin{abstract}
Various neural network-based approaches have been proposed for more robust and accurate voice activity detection (VAD).
Manual design of such neural architectures is an error-prone and time-consuming process, which prompted the development of neural architecture search (NAS) that automatically design and optimize network architectures.
While NAS has been successfully applied to improve performance in a variety of tasks, it has not yet been exploited in the VAD domain.
In this paper, we present the first work that utilizes NAS approaches on the VAD task.
To effectively search architectures for the VAD task, we propose a modified macro structure and a new search space with a much broader range of operations that includes attention operations.
The results show that the network structures found by the propose NAS framework outperform previous manually designed state-of-the-art VAD models in various noise-added and real-world-recorded datasets.
We also show that the architectures searched on a particular dataset achieve improved generalization performance on unseen audio datasets.
Our code and models are available at \href{https://github.com/daniel03c1/NAS\_VAD}{https://github.com/daniel03c1/NAS\_VAD}.

\end{abstract}


\section{Introduction}
\label{sec:intro}
Voice activity detection (VAD) is a task of detecting the existence of human voice and its onset and offset times.
Early studies have utilized feature engineering and statistical signal processing methods to detect voice activity from audio signals~\cite{SohnsVAD,rVAD}.
The adoption of deep learning approaches has demonstrated a significant improvement in detection performance, especially in noisy environments~\cite{DBN,zhang2015boosting,hughes2013recurrent}.
In order to further improve detection performance at low signal-to-noise ratios (SNRs), several studies have exploited contextual information in various ways and demonstrated improved performance~\cite{zhang2015boosting,kim2018voice,ST-Attention,SA}.
Zhang et al.~\cite{zhang2015boosting}, for example, proposed aggregating predictions from different temporal contexts, and other approaches employed variations of attention mechanisms to leverage contextual information~\cite{kim2018voice, ST-Attention, SA}.
While the existing deep learning based VAD approaches have shown remarkable progress, there is still room for further optimization of the neural architectures to achieve higher VAD accuracy with smaller model size.
However, manual design of neural architecture requires significant engineering efforts of human designers, which becomes increasingly more challenging and time-consuming as the architecture gets larger and more complicated~\cite{nas_survey,naswot}.

The need for a reduction of human effort in network design has led to the emergence of neural architecture search (NAS) that automatically optimizes model structures and operations.
Some of the NAS methods have succeeded in creating models that outperform manually-designed models in image classification \cite{Zoph_2018_CVPR, design_space} and automatic speech recognition tasks \cite{li20n_interspeech,Chen2020DARTS-ASR, Kim2020EvolvedSpeechTransformer}.
Despite the successful adoption of NAS for various tasks, no research has explored applying NAS to the VAD task.

In this work, we propose to adopt an architecture search mechanism on the VAD task to facilitate detection performance improvement through automatic network architecture design.
To the best of our knowledge, this is the first attempt to apply NAS to the VAD task.
Based on a cell-based NAS approach, we propose expanding the search space to include the attention mechanism and its related operations known to be effective for the VAD task~\cite{ST-Attention,SA}, and modifying macro-architecture to reflect shallow depths of VAD models. 
The experimental results show that NAS can discover VAD models that outperform the existing manually-designed network structures across a variety of audio datasets with a restricted number of parameters.
In addition, the models searched by the proposed framework demonstrate enhanced generalization performance, which is an important feature of the VAD task.

\section{Background and Related Work}
\label{sec:format}

In the field of deep learning, neural architecture search has become an important technique that can greatly reduce manual effort in neural network design.
Models optimized by NAS have shown comparable or even better performance than manually-designed architectures on various computer vision tasks~\cite{Zoph_2018_CVPR, pham2018efficient, design_space}.
In terms of the search strategy, several lines of strategies have been proposed, including evolutionary methods \cite{real2017largeevoluation}, Bayesian optimization \cite{zhou2019bayesnas,ru2021interpretable}, reinforcement learning \cite{zoph2017,pham2018efficient}, gradient-based methods \cite{liu2018darts}, and random search \cite{li2020randomsearch}.
The scope of search space is another consideration for architecture search.
The search space can be the whole architecture for a given set of input and output shapes~\cite{real2017largeevoluation}, or only the inner structure (operations and connections) for a fixed macro architecture~\cite{design_space}, which acts as an exoskeleton and is not optimized during architecture search.
A cell-based search space further reduces the search space by filling the inner structure with a repeating pattern, ``cell''~\cite{Zoph_2018_CVPR, pham2018efficient}.
As for the cell-based search space, a neural network is defined as a stack of one or two types of cells, and only the operations and connections of each cell type are optimized during architecture search.
Cell-based search spaces are known to be more effective than much larger search spaces since they require fewer model evaluations to find optimal architectures without compromising performance~\cite{Zoph_2018_CVPR}.

NAS is being widely adopted in a variety of vision and natural language processing tasks~\cite{li20n_interspeech}.
Recently, audio processing tasks such as acoustic scene classification \cite{li20n_interspeech} and automatic speech recognition \cite{Chen2020DARTS-ASR,Kim2020EvolvedSpeechTransformer} have begun to utilize NAS to improve model performance and design efficiency.
In these speech-related tasks, several studies adjusted the search space~\cite{li20n_interspeech} or macro structure~\cite{Ding2020AutoSpeechNA} for speech processing, but they did not include sequential operations, which have been shown to improve VAD performance.
Furthermore, there has been no study that explored the application of NAS to the VAD task.
In this paper, we demonstrate the efficacy of NAS in optimizing neural architecture for enhanced voice detection performance, using a newly expanded search space and a modified macro structure.

\begin{figure}
\begin{center}
\includegraphics[width=0.625\linewidth]{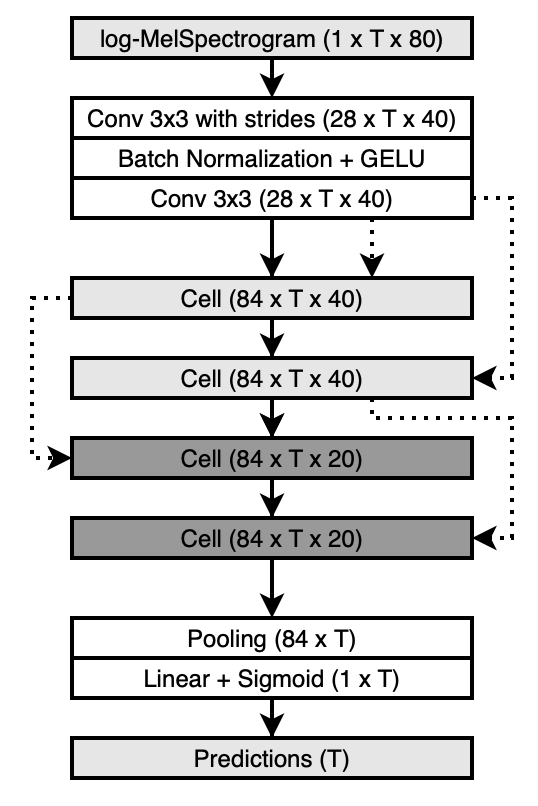} 
\end{center}
\caption{The proposed macro structure. Each block denotes an operation or a module, and the parenthesis beside it shows the output shape of it. The first dimension is the channel dimension, and the second dimension, denoted as $T$, represents the time dimension, and the third number indicates the feature dimension. Dashed lines denote the second connection for each cell.}
\label{fig:subim1}
\end{figure}

\section{NAS-VAD}
\label{sec:vad-nas}

In this paper, we employ NAS for the VAD task to automatically design the network architecture that can yield improved detection performance.
For computational efficiency, we adopt a Bayesian optimization-based search strategy~\cite{ru2021interpretable} on a cell-based search space.
Our proposed search space includes diverse operations, including those that have been shown to be beneficial for the VAD task.
We also modified the macro structure in order to discover models with lesser depths and better performance. 

\subsection{Search Strategy}
\label{ssec:search_strategy}

To search for the optimal VAD model architecture, we use the cell-based Bayesian optimization approach~\cite{ru2021interpretable}, which alleviates the heavy cost of training and evaluating models using the Gaussian process as a surrogate model.
Among Bayesian optimization based search methods, we adopt NAS-BOWL~\cite{ru2021interpretable}, which allows sample-efficient search compared to other Bayesian optimization based methods through the Weisfeiler-Lehman (WL) subtree graph kernel.

\subsection{Search Space}
\label{ssec:search_space}

We propose a new search space for the VAD task because the default search space used in NAS-BOWL was designed for image classification tasks and lacks important operations for the VAD task, including attention operations.
To exploit the contextual information, we include the attention operation and its related operations (e.g., feed forward network modules)~\cite{vaswani2017attention} into the search space.
We adopt a total of 18 operations into the search space.
Among them, four operations are convolutional operations, and the other six operations are attention operations.
In addition, three feed forward networks~\cite{vaswani2017attention}, one squeeze-and-excitation operation~\cite{hu2018squeeze}, two gated-linear units (GLU)~\cite{dauphin2017language}, skip connection, and zero operation (to represent a disconnection) have also been adopted.

As for convolutional operations, we used four bottleneck residual block types~\cite{sandler2018mobilenetv2}.
This includes 3x3 and 5x5 bottleneck residual blocks, each of which can have an expansion rate of 2 and 4.
For bottleneck residual blocks, we use GELU \cite{hendrycks2016gaussian} as an activation function.

We used three variations of multi-head attention~\cite{vaswani2017attention}: one that just applies the attention mechanism on time, another on feature, and the third on both time and feature dimensions.
This was motivated by the spectro-temporal attention VAD~\cite{ST-Attention}, which improved detection performance by using two types of attention operations: one that only applies in time and the other that only applies on feature dimensions.
The number of heads for each attention type was set at two or four, which resulted in a total of six attention types.

In addition, feed forward networks (FFN) \cite{vaswani2017attention} with an expansion rate of 0.5, 1, and 2, and gated linear units \cite{dauphin2017language} with a kernel size of 3x3 and 5x5 have been adopted.
We adopted pre-activation \cite{he2016identity} in bottleneck residual blocks, multi-head attention layers, and feed-forward networks. 

\subsection{Macro Architecture}
\label{ssec:macro_arc}


Our newly proposed macro architecture for NAS-VAD is shown in Fig.~\ref{fig:subim1}.
As opposed to the basic structure~\cite{ru2021interpretable}, we use only four cells because general VAD models~\cite{BDNN, kim2018voice, ST-Attention, SA} have a shallower depth for lower latency.
Unless the input features are already in the expected shape, average pooling is applied to the input features to match the expected feature dimension size, followed by a convolutional layer with a 1x1 kernel to match the number of channels for each input port of the cells.
Feature reductions were applied on cells located at 1/2 of the total depth (the number of cells).
In the original reduction stage used in NAS-BOWL, both height (time) and width (features) are halved, and the number of channels is doubled.
However, because we need to keep the size of time dimension intact, we only reduce the width and not height.
We also employ two convolutional layers with batch normalization and GELU activation in front of the first cell, and add a linear layer right after the last cell output to generate the predictions.

\begin{table*}[ht]
\caption{Comparison of model performance}
\label{table_2}
\centering
\begin{threeparttable}
\begin{tabular}{l|l|l|cccc} 
\toprule
\multicolumn{1}{l}{} & \multicolumn{1}{l}{} & \multicolumn{1}{l}{} & \multicolumn{2}{c}{TIMIT + SoundIdeas} &   \multicolumn{2}{c}{CV + Audioset} \\ 
\cmidrule(r){4-5} \cmidrule(r){6-7} 
\multicolumn{1}{l}{trainset} & \multicolumn{1}{l}{model} & \multicolumn{1}{l}{\# params} &\multicolumn{1}{c}{AUC}  &\multicolumn{1}{c}{F1-score}  & \multicolumn{1}{c}{AUC} & \multicolumn{1}{c}{F1-score} \\
\midrule
& NAS-VAD & 151k & \textbf{0.982$\pm$0.003}  & \textbf{0.926$\pm$0.007} & \textbf{0.824$\pm$0.007} & \textbf{0.748$\pm$0.009} \\
\cmidrule(r){2-7} 
TIMIT & bDNN~\cite{zhang2015boosting}& 498K & 0.968$\pm$0.001 & 0.899$\pm$0.003 & 0.747$\pm$0.003 & 0.709$\pm$0.003 \\
 + SoundIdeas & ACAM~\cite{kim2018voice} & 342K & 0.960$\pm$0.009 & 0.885$\pm$0.015 & 0.717$\pm$0.016 & 0.700$\pm$0.004 \\
& Spectro-temporal attention~\cite{ST-Attention} & 558K & 0.977$\pm$0.003 & 0.915$\pm$0.007 & 0.781$\pm$0.017 & 0.724$\pm$0.007 \\ 
& Self-attentive VAD~\cite{SA} & 155K & 0.974$\pm$0.003  & 0.910$\pm$0.006 & 0.797$\pm$0.013 & 0.737$\pm$0.011 \\
\midrule
& NAS-VAD & 151k & \textbf{0.965$\pm$0.001} & \textbf{0.894$\pm$0.001} & \textbf{0.976$\pm$0.001} & \textbf{0.915$\pm$0.002} \\
\cmidrule(r){2-7} 
CV + Audioset & bDNN~\cite{zhang2015boosting}& 498K & 0.941$\pm$0.000 & 0.861$\pm$0.002 & 0.950$\pm$0.000 & 0.871$\pm$0.000 \\ 
& ACAM~\cite{kim2018voice} & 342K & 0.940$\pm$0.002 & 0.860$\pm$0.003 & 0.948$\pm$0.002 & 0.869$\pm$0.003 \\ 
& Spectro-temporal attention~\cite{ST-Attention} & 558K & 0.955$\pm$0.001 & 0.879$\pm$0.001 & 0.968$\pm$0.001 & 0.894$\pm$0.012 \\ 
& Self-attentive VAD~\cite{SA} & 155K & 0.951$\pm$0.001 & 0.873$\pm$0.001 & 0.963$\pm$0.001 & 0.890$\pm$0.002 \\ 
\bottomrule
\end{tabular}
\vspace{0.3ex}
\end{threeparttable}
\end{table*}

\begin{table}
\caption{Detection performance of VAD models on real-world datasets (AUC)}
\label{table:real_world}
\centering
\resizebox{\columnwidth}{!}{\begin{tabular}{l|l|ccccc}
\toprule
\multicolumn{1}{l}{} & \multicolumn{1}{l}{} & \multicolumn{5}{c}{Model} \\
\cmidrule(r){3-7}
\multicolumn{1}{l}{trainset} & \multicolumn{1}{l}{testset} & \multicolumn{1}{c}{Ours}  &\multicolumn{1}{c}{bDNN}  & \multicolumn{1}{c}{ACAM} & \multicolumn{1}{c}{STA$^a$} & \multicolumn{1}{c}{SA$^b$} \\
\midrule
      & AVA    & 0.774 & 0.787 & 0.792 & 0.789 & \textbf{0.836} \\
\cmidrule(r){2-7}
      & bus stop$^c$ & \textbf{0.980} & 0.876 & 0.924 & 0.962 & 0.954 \\
TIMIT & cons. site$^c$ & \textbf{0.990} & 0.970 & 0.965 & 0.985 & 0.985 \\
      & park$^c$ & \textbf{0.991} & 0.989 & 0.982 & 0.990 & 0.990 \\
      & room$^c$ & 0.980 & 0.983 & 0.984 & 0.983 & \textbf{0.988} \\
\midrule
   & AVA    & \textbf{0.905} & 0.868 & 0.858 & 0.892 & 0.892 \\
\cmidrule(r){2-7}
   & bus stop$^c$ & \textbf{0.989} & 0.987 & 0.987 & 0.986 & 0.987 \\
CV & cons. site$^c$ & \textbf{0.998} & 0.996 & 0.996 & 0.997 & 0.997 \\
   & park$^c$ & \textbf{0.991} & 0.989 & 0.989 & 0.990 & 0.989 \\
   & room$^c$ & \textbf{0.994} & 0.993 & 0.992 & 0.993 & 0.993 \\
\bottomrule
\end{tabular}}
\begin{tablenotes}
\item $^a$ spectro-temporal attention VAD, $^b$ self-attentive VAD, \\
$^c$ real-world audioset from ACAM.
\end{tablenotes}
\end{table}
\vspace*{5mm}

\begin{figure}[ht]
    \centering
    \includegraphics[width=0.9\linewidth]{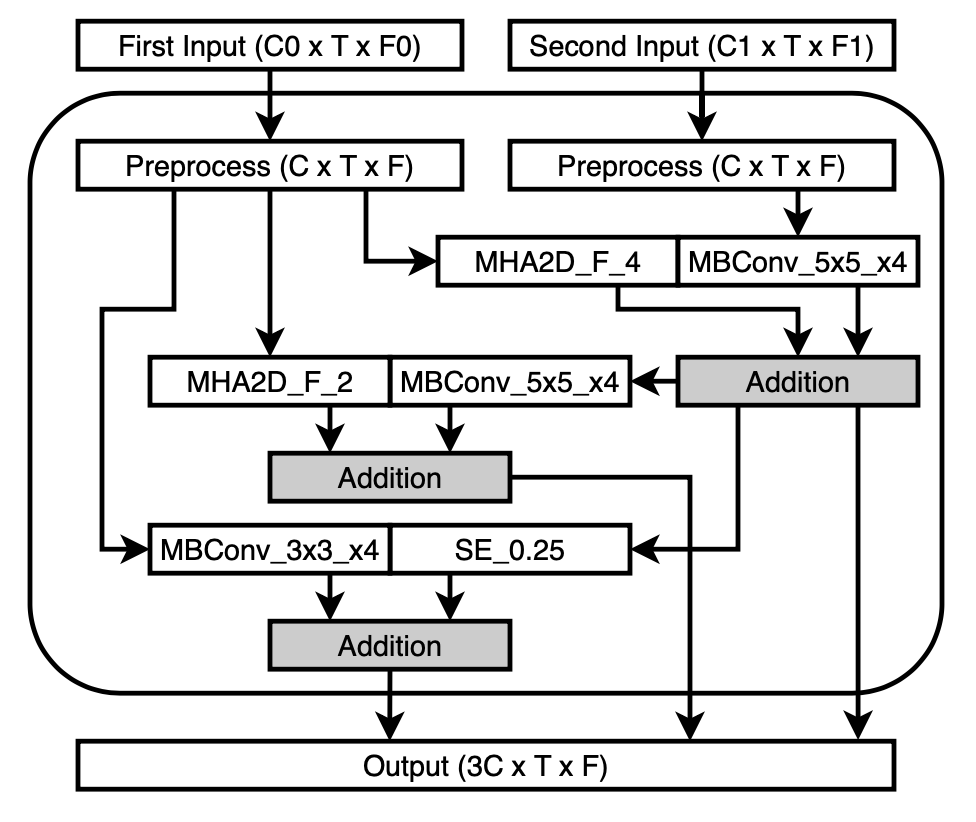}
    \captionsetup{justification=justified, singlelinecheck=off}
    \caption{The cell discovered on TIMIT + SoundIdeas within the proposed search space. The size of the channel, time, and feature dimension are denoted by the letters C, T, and F, respectively. \emph{Preprocess} is a function to map input features to the appropriate feature space. If the input channels and the number of features are as expected, it will simply pass on to the next operation. Otherwise, it will be preprocessed as described in Sec.~\ref{ssec:macro_arc}. \emph{MBConv\_3x3\_x4} and \emph{MBConv\_5x5\_x4} are MobileNetV2 bottleneck residual blocks with an expansion rate of 4 and different kernel sizes (3x3 and 5x5). \emph{MHA2D\_F\_2} and \emph{MHA2D\_F\_4} are multi-head attention but only apply to the feature dimension, which has two and four heads, respectively. \emph{SE\_0.25} is a squeeze-and-excitation module with a squeeze ratio of 0.25. Lastly, the three outputs resulting from the \emph{addition} nodes are concatenated to construct a final result.}
    \label{fig:cell}
\vspace*{-5mm}
\end{figure}

\section{Experiments}
\label{sec:experiments}

To evaluate the effectiveness of our NAS-VAD framework, we compared the detection performance and the parameter size of the model searched by NAS-VAD with other baseline models.
We used manually designed, state-of-the-art VAD models as baselines.
These baselines are built for a frame-wise voice detection framework using non-consecutive audio frames.
As a result, we employed the same VAD framework for a fair comparison.
Because stochastic and algorithmic VAD approaches could not work on these non-consecutive audio frames, we did not compare these models with neural network-based models.

\subsection{Experiment Settings}
\label{ssec:experiment_setups}

\subsubsection{Datasets}
\label{ssec:exp_dataset}

We used noise-added speech datasets for both architecture search and model evaluation, and real-world datasets solely for performance evaluation.
We used TIMIT~\cite{TIMIT} and Common Voice 7.0~\cite{Ardila2020CommonVA} as clean voice sources and added noises from SoundIdeas~\cite{soundIdea} and Audiosets~\cite{audioset}, respectively, to create noise-added speech datasets.
We used the entire audio samples from the TIMIT dataset, which contains approximately 5.4 hours of English utterances.
The CV dataset~\cite{Ardila2020CommonVA} is a publicly available multi-lingual voice dataset of more than 10,000 hours of utterances.
We only used 2\% of the total audio samples in the CV dataset, which is already 40 times longer than the TIMIT dataset.
Because the CV dataset lacks labels for the VAD task, we used an algorithmic VAD method to extract ground truth labels from clean audio samples of the CV dataset, following VAD research conventions when ground truth labels are lacking.
We used the entire SoundIdeas audio sample collection, which offers a wide range of non-speech noises such as wind and short sound effects such as a dog barking.
Audioset is a publicly available, human-labeled audio dataset extracted from YouTube.
We only used selected non-speech noises preprocessed by Reddy et al.\cite{reddy2021interspeech} because the Audioset contains many audio samples containing human voices.
Before adding noises, we padded each clean speech audio sample in order to make the number of speech and non-speech frames equal.
After that, we added a randomly chosen noise to the padded speech audio sample.
The SNR of noise-added audio samples was uniformly sampled from -10 to 10 dB, and SNR was calculated based on active regions of a clean speech sample and a noise audio sample.
To measure the robustness of VAD models in real-world recordings, we used the AVA~\cite{AVA} and ACAM datasets~\cite{kim2018voice} as test datasets.

We divided each audio and noise dataset into training, validation, and test groups at a ratio of 8:1:1 unless otherwise specified by the dataset.
As an input feature, we used log-melspectrogram with 80 mel filterbanks, viewing each spectrogram as a single channel image.
To compare with earlier VAD networks~\cite{zhang2015boosting, kim2018voice, ST-Attention, SA}, we trained each model to detect voice activity in seven neighboring frames [t-19, t-10, t-1, t, t+1, t+10, t+19] for a given frame at time t.

\subsubsection{Architecture search}
We used the TIMIT-based dataset to search for an architecture.
As a metric, we used Area Under the Curve (AUC), which is a commonly used metric in the VAD task performance evaluation.
The default number of channels for neural networks was set to 16 in order to find VAD models with the model size similar to that of the smallest state-of-the-art VAD model~\cite{SA}. 

\subsubsection{Model Evaluation}
For performance comparison, we used the following manually-designed models: bDNN \cite{BDNN}, ACAM~\cite{kim2018voice}, spectro-temporal attention VAD \cite{ST-Attention}, and self-attentive VAD \cite{SA}.
For self-attentive VAD, we used the official model implementation, and for other models, we implemented them as described in the original papers.
To train VAD models, including ours, we set the maximum training epochs to 50 for CV and 100 for TIMIT.
We used the Adam optimizer with an initial learning rate of 1e-3 and the cosine annealing learning rate scheduler~\cite{Cosineschedule}.
We used a batch size of 128 and early-stopped training if the loss did not decrease for ten consecutive epochs.
For data augmentation, we used random volume scaling and frequency masking~\cite{SpecAugment}.
As in the baseline VAD models~\cite{zhang2015boosting, kim2018voice}, we used boosted frame-wise predictions when evaluating each model to improve voice detection performance.
As for detection performance metrics, we used both AUC and f1-score, as commonly used in previous VAD studies~\cite{zhang2015boosting, kim2018voice,chen20o_interspeech}.
We aggregated ground truth labels and corresponding predictions across multiple files to measure AUC.
We binarized outputs with a threshold of 0.5 and used these binarized predictions to measure the F1-score of each model.
We trained each model at least three times and reported the average test performance and its standard deviation.

To evaluate the detection performance of VAD models, we first trained and tested them on two noise-added speech datasets.
After that, we also measured the performance on real-world datasets to compare the generalization performance on unseen real-world recorded audio samples.

\subsection{Results}
\label{sec:results}

The cell structure learned on the TIMIT-based dataset in our proposed search space is depicted in Fig. \ref{fig:cell}.
Table~\ref{table_2} shows the voice detection performance of our model and four baseline models on noise-added audio datasets.
The model from our proposed search space and macro architecture shows significant performance improvement compared to the manually-designed models, outperforming every other VAD model regardless of the training and test datasets.
Even though we ran architecture searches only on the TIMIT-based dataset, the resulting model showed improved detection performance also in the CV-based dataset.
Furthermore, the performance gap between ours and other VAD models is markedly widened when trained on the CV-based dataset.
This demonstrates that, at least in our experiments, searching for architectures in a particular dataset does not necessarily result in a decrease in generalization performance.
Despite enhanced detection performance, our model has smaller number of parameters (151K) than any other network, as demonstrated in Table~\ref{table_2}.

Table~\ref{table:real_world} shows the generalization performance of various VAD models on unseen real-world audio samples.
Our model outperforms competing VAD models, achieving the best detection performance across most test datasets, as it did in the noise-added audio dataset.
In particular, when trained on the CV-based dataset, our model outperformed on every real-world dataset.
When trained on the TIMIT-based dataset, our model failed to outperform self-attentive VAD in room dataset~\cite{kim2018voice} and in AVA.
We believe this is due to the fact that our model can learn important features better than existing VAD models, despite having a smaller number of parameters, resulting in overfitting to noise-added datasets when trained on a small number of noise-added, artificial audio samples.

In summary, using NAS, we have succeeded in finding VAD models that outperform other state-of-the-art VAD models while maintaining a similar number of parameters to other VAD models.

\subsection{Ablation Study}
\label{ssec:ablation}

We ran an ablation study to measure the performance improvement brought by the newly proposed search space and the macro architecture.
We trained each model, one from the original DARTS search space and the other from our search space, on the TIMIT-based dataset and tested them on two noise-added audio datasets.
As shown in  Table~\ref{table:ablation}, our proposed search space and macro architecture improved performance, especially in the CV dataset.
Considering that we used the same search algorithm and that the resulted networks have a similar number of parameters, the performance gap can only be attributed to the modification of search space and macro structure.
This demonstrates the efficacy of expanding search space by incorporating various operations, such as attention operations, along with macro structure modification.

\begin{table}
\caption{Search space comparison}
\label{table:ablation}
\centering
\resizebox{\columnwidth}{!}{
\begin{threeparttable}
\begin{tabular}{l|l|cccc}
\toprule
\multicolumn{1}{l}{} & \multicolumn{1}{l}{} & \multicolumn{2}{c}{TIMIT} &   \multicolumn{2}{c}{CV} \\ 
\cmidrule(r){3-4} \cmidrule(r){5-6} 
\multicolumn{1}{l}{search space} & \multicolumn{1}{l}{\# params} &\multicolumn{1}{c}{AUC}  &\multicolumn{1}{c}{F1}  & \multicolumn{1}{c}{AUC} & \multicolumn{1}{c}{F1} \\
\midrule
Ours & 151k & \textbf{0.982}  & \textbf{0.926} & \textbf{0.824} & \textbf{0.748} \\
DARTS & 169K & 0.981 & 0.923 & 0.807 & 0.732 \\
\bottomrule
\end{tabular}
\vspace{0.3ex}
\end{threeparttable}}
\vspace{-5.5mm}
\end{table}

\section{Conclusion}
\label{sec:conclusion}
To improve VAD performance through an automatic design process, we proposed a NAS framework with search space and macro-structure optimized for the VAD task.
The results demonstrate that the models searched by NAS-VAD outperform the existing VAD models.
We believe developing neural network-based VAD models in this framework might provide inspiration for more robust and accurate VAD models.
We expect that the proposed framework can also be used to improve other audio processing tasks as well.

\bibliographystyle{IEEEtran}
\bibliography{template}
\end{document}